\documentclass[10pt,a4paper,twoside]{article}
\usepackage{epsfig}
\usepackage{baltlat6}
\usepackage{array}
\usepackage{longtable}
\usepackage{wrapfig}
\begin{document}
\ \ \vspace{0.5mm} \setcounter{page}{1} \vspace{8mm}

\titlehead{Baltic Astronomy, vol.\,20, 00--00, 2011}

\titleb{THE LOWER MAIN SEQUENCE OF STARS IN THE SOLAR
NEIGHBORHOOD: MODEL PREDICTIONS VERSUS\\ OBSERVATION}

\begin{authorl}
\authorb{S. Barta\v{s}i\={u}t\.e}{},
\authorb{V. Deveikis}{},
\authorb{S. Raudeli\={u}nas}{}
and
\authorb{J.~Sperauskas}{}
\end{authorl}

\begin{addressl}

\addressb{}{Astronomical Observatory of Vilnius University,
\v{C}iurlionio 29, Vilnius,\\ LT-03100, Lithuania}

\end{addressl}

\submitb{Received: 2011 September 15; accepted 2011 October }

\begin{summary} We have used the Simbad database and VizieR catalogue
access tools to construct the observational color--absolute
magnitude diagrams of nearby K--M dwarfs with precise {\it
Hipparcos} parallaxes ($\sigma_\pi/\pi\leq 0.05$). Particular
attention has been paid to removing unresolved double/multiple stars
and variables. In addition to archival data, we have made use of
nearly 2000 new radial-velocity measurements of K--M dwarfs to
identify spectroscopic binary candidates. The main sequences,
cleaned from unresolved binaries, variable stars, and old population
stars which can also widen the sequence due to their presumably
lower metallicity, were compared to available solar-metallicity
models. Significant offsets of most of the model main-sequence lines
are seen with respect to observational data, especially for the
lower-mass stars. Only the location and slope of the Victoria-Regina
and, partly, BaSTI isochrones match the data quite well.
\end{summary}


\resthead{The lower main sequence stars}{S. Barta\v{s}i\={u}t\.e, V.
Deveikis, S. Raudeli\={u}nas, J. Sperauskas}

\sectionb{1}{INTRODUCTION}

The validation of stellar theoretical models relies heavily upon the
accurate determination of the location and properties of the
sequences of stars in the color--luminosity diagrams. For
main-sequence stars with masses above solar, most of the theoretical
models are generally in line with observations. The situation is
much worse when we discuss the domain of low-mass ($M<0.8\,M_\odot$)
stars. On one hand, the complexity of their spectra with molecular
features dominating complicates both the calculation of stellar
models and the color-temperature transformations, what leads to
difficulties in determining good theoretical main-sequence lines. On
the other hand, comparisons of models to data require comprehensive
observations from which it would be possible to accurately locate
the observational ridge lines of different metallicity. Because of
the inefficiency of photometric methods to determine metallicities
of M-type dwarfs, the local samples suitable for such comparisons
are too modest in size. Therefore, the more generally accepted
approach has been to use instead of field stars the Galactic star
clusters. With this approach, however, the advantage in having the
lower main sequence of homogeneous chemical composition is often
reduced by uncertainties coming from fitting procedure (distances,
etc.) and the scatter within the fainter portion of the sequence due
to observational limitations.

In recent years, large photometric surveys such as the Sloan Digital
Sky Survey (SDSS) and the Two Micron All Sky Survey (2MASS) have
proved to be superb data resources for statistical investigations of
late-type dwarfs. However, accurate parallaxes are not available for
the overwhelming majority of faint stars that could probe the
lower-main sequence. Thus, until {\it Gaia} space mission becomes a
reality, nearby {\it Hipparcos} stars still remain demanding
objects.

In the paper by Kotoneva et al. (2002), {\it Hipparcos}-based
absolute magnitudes of K dwarfs were compared with a set of
theoretical isochrones for three metallicity ranges -- solar,
subsolar ([Fe/H] between --\,0.30 and --\,0.50) and super-solar
([Fe/H]=0.18\,--\,0.30). The metallicities were derived by
photometric method using {\it Geneva} $b_1$ and Johnson-Cousins
$(R-I)_{\rm C}$ or Str{\"o}mgren $m_1$ and $(R-I)_{\rm C}$ colors.
They found a tight relationship between luminosity, color and
metallicity for K dwarfs. However, none of the isochrones tested by
them in the $M_V$,$B$--$V$ plane fitted the observational K-dwarf
sequences.

Just \& Jahrei{\ss} (2008) investigated the properties of the main
sequence of {\it Hipparcos} F--K stars in the Catalogue of Nearby
Stars (CNS), using the Johnson-Cousins $BV(RI)_{\rm C}$ data
transformed to the SDSS $ugriz$ filter system. Systematic
differences in the shape and location of the main sequences in the
$ugriz$ system were found with respect to the theoretical isochrones
of Padova and Dartmouth models. More recently, Bochanski et al.
(2010) have derived color -- absolute magnitude relations for
low-mass stars using new yet unpublished $(ugriz)^\prime$
observations and 2MASS $JHK_{\rm S}$ photometry of nearby dwarfs
with known trigonometric parallax measurements. These relations have
been used by the SDSS team to estimate absolute magnitudes and
distances to all faint stars in their huge sample.

The aim of this paper is to make a comparison of the sequence of
nearby K--M dwarfs in the ($M_V$, color) diagrams, based on the data
available at the CDS, with theoretical solar-metallicity isochrones
of various stellar models. We accept that the ridge lines, defined
in the ($M_V$, color) diagram by stars with kinematics typical of
the young to intermediate age disk, may adequately represent the
locus of solar-metallicity stars. The absolute magnitudes of K--M
dwarfs will be solely based on {\it Hipparcos} parallaxes. Our
analysis in this paper is restricted to Johnson-Cousins $BV(RI)_{\rm
C}$ photometry as in this particular system we can find the majority
of accurate observational data on nearby stars and the majority of
model isochrones calculated. We focus attention on the main
contributors to the scatter of the observational sequence --
photometric variability, unresolved binarity and measurement errors.
While for information on variability and astrometric multiplicity of
nearby stars we can address the {\it Hipparcos} survey and expanding
datasets of supplementary observations, the census of spectroscopic
binaries, which requires long-term radial-velocity programs, is far
from being complete. Therefore, to identify new unresolved binaries
among nearby K--M dwarfs, we have made use of nearly 2000
radial-velocity measurements obtained over the past decade within
our CORAVEL program (Upgren, Sperauskas \& Boyle 2002).

In the following section we will describe the selection of stars,
refining of the samples and observational data used. Comparisons of
the data to theoretical isochrones of different stellar models are
demonstrated in Section 3.

 \begin{figure}[!t]
\centerline{\psfig{figure=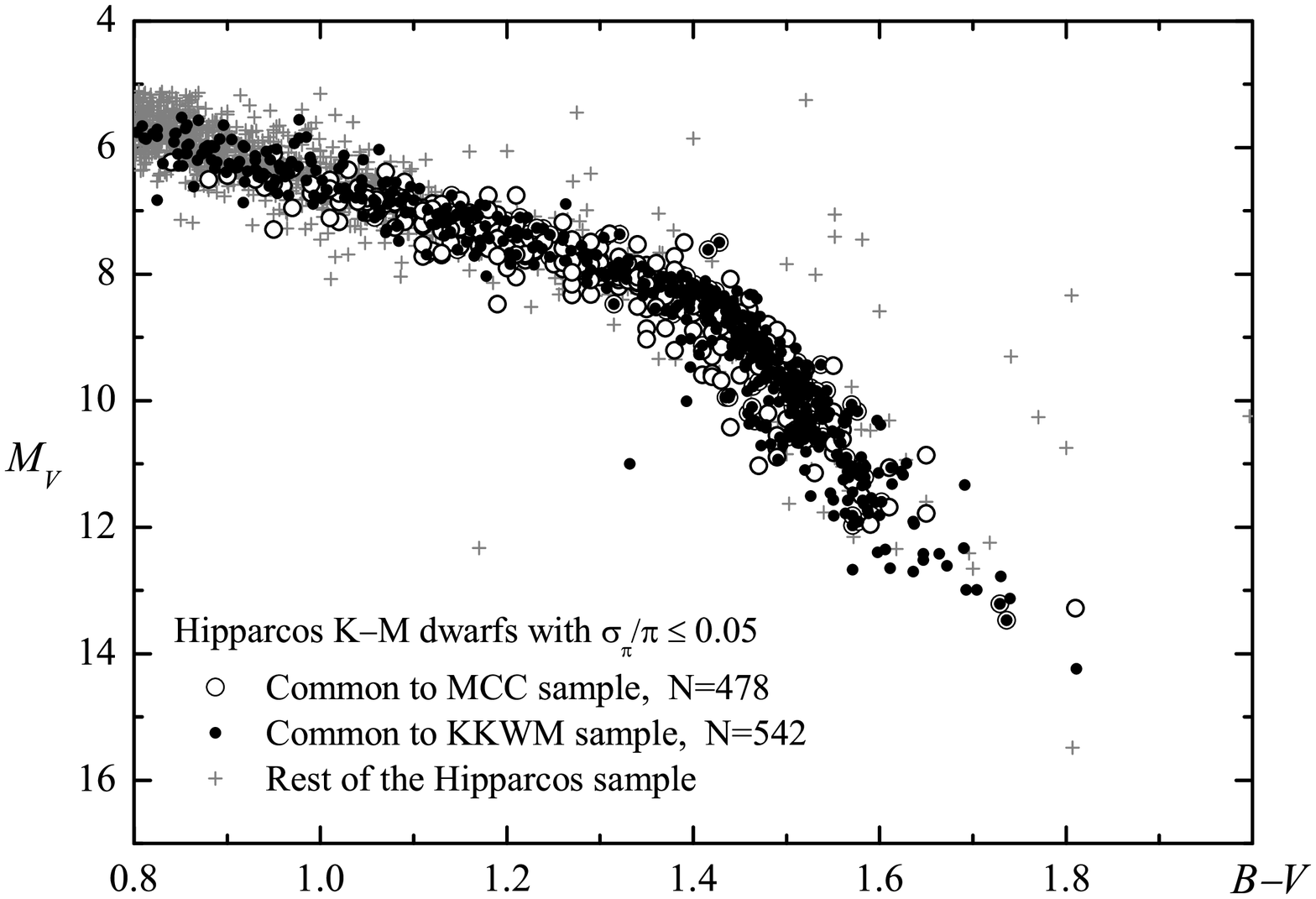,width=90mm,angle=0,clip=}}
\captionb{1}{$M_V$, $B$--$V$ diagram for {\it Hipparcos} K--M dwarfs
with $\sigma_\pi/\pi\leq0.05$. $BV$ photometry to plot MCC stars
(open circles) and KKWM stars (solid points) is from CNS3 and Koen
et al. (2010), respectively, while the rest of the stars (grey
crosses) are plotted using $BV$ data taken from the {\it Hipparcos}
main catalog.}
\end{figure}

\sectionb{2}{DATA AND SAMPLE SELECTION}

We started with the selection of stars with precise trigonometric
parallaxes from {\it Hipparcos} catalog (van Leeuwen 2007), which
fall within the typical color and luminosity range for K and M
dwarfs: $B$--$V$\,$>$\,0.80 and $M_V \geq 5.5$ mag. A limit on
parallax accuracy of $\sigma_\pi/\pi\leq 0.05$ was chosen to ensure
the absolute magnitudes to be accurate to within $\sim 0.1$ mag,
with a negligible bias introduced by the Lutz-Kelker effect. We have
1815 stars that satisfy the above criteria (Figure~1). However, for
nearly half of these stars, no homogeneous and accurate photometry
or supplementary data can be found in the existing databases. To
have the luminosities and colors based on rigorous observational
basis, we have finally chosen for further analysis two separate
samples of {\it Hipparcos} K--M dwarfs:
\begin{itemize}
    \item[(1)]The McCormick sample (hereafter MCC), which constitutes stars
north of declination $-30^\circ$, selected years ago
spectroscopically at the McCormick Observatory by A.\,N.~Vyssotsky
and his colleagues (for a review of their survey, see Upgren \& Weis
1989). For all of these stars, the $BV(RI)_{\rm C}$ data are
thoroughly collected at the ARI Database for Nearby Stars (CNS3,
Gliese \& Jahreiss (1991), with its more updated version of 1998).
    \item[(2)]The sample of stars south of declination $+26^\circ$
    from the recently published catalog
of homogeneous and standardized $UBV(RI)_{\rm C}JHK$ photometry by
Koen, Kilkenny, van Wyk \& Marang (2010) (hereafter KKWM).
    \end{itemize}
The coverage of color -- luminosity diagram by the MCC and KKWM
stars with respect to the remaining {\it Hipparcos} stars with
$\sigma_\pi/\pi\leq 0.05$ is shown in Figure~1. There was no need to
apply the extinction and reddening corrections, since the stars in
both samples are at distances smaller than 50 pc (the KKWM  stars
are within 30 pc).

All stars in the MCC sample and part of northernmost stars in the
KKWM sample, which had no high-quality or any radial-velocity
measurements, were targets of our decadal program of radial-velocity
observations (Upgren, Sperauskas \& Boyle 2002). However, 126 stars
of the selected KKWM sample are still lacking good radial-velocity
measurements and thus were rejected from further analysis. Thus we
have 478 stars in the MCC sample and 416 stars in the KKWM sample,
which satisfy the $\sigma_\pi/\pi\leq 0.05$ criterion and have
radial-velocity data of satisfactory quality. Of these, 173 stars
are found to be common to both samples.

 \begin{figure}[t!]
\centerline{\psfig{figure=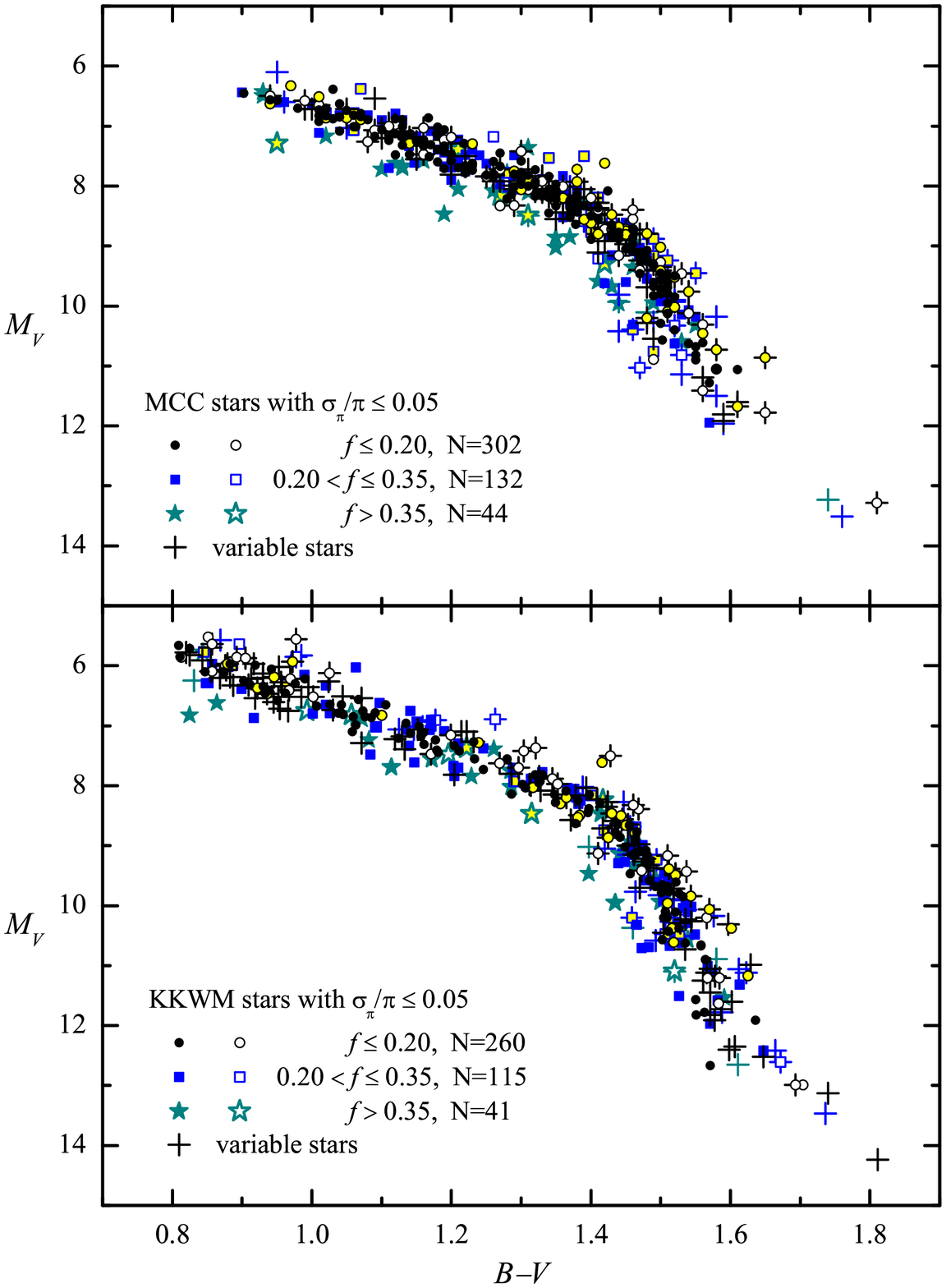,width=90mm,angle=0,clip=}}
\captionb{2}{$M_V$, $B$--$V$ diagrams for K--M dwarfs with
$\sigma_\pi/\pi\leq 0.05$ in the two samples. Different symbols
denote different kinematic groups: young to intermediate age thin
disk ($f\leq 0.20$; circles), old thin disk ($0.20<f\leq0.35$;
squares) and thick disk ($f>0.35$; five-sided stars). Stars showing
no indication of variability or multiplicity are indicated by filled
symbols, while open symbols denote known and suspected binaries (in
color version, the symbols for spectroscopic binaries and
radial-velocity variables are filled with yellow). Variable stars
are shown by crosses (if the star is both variable and binary, a
cross is overlaid by an open symbol).}
\end{figure}

 \begin{figure}[t!]
\centerline{\psfig{figure=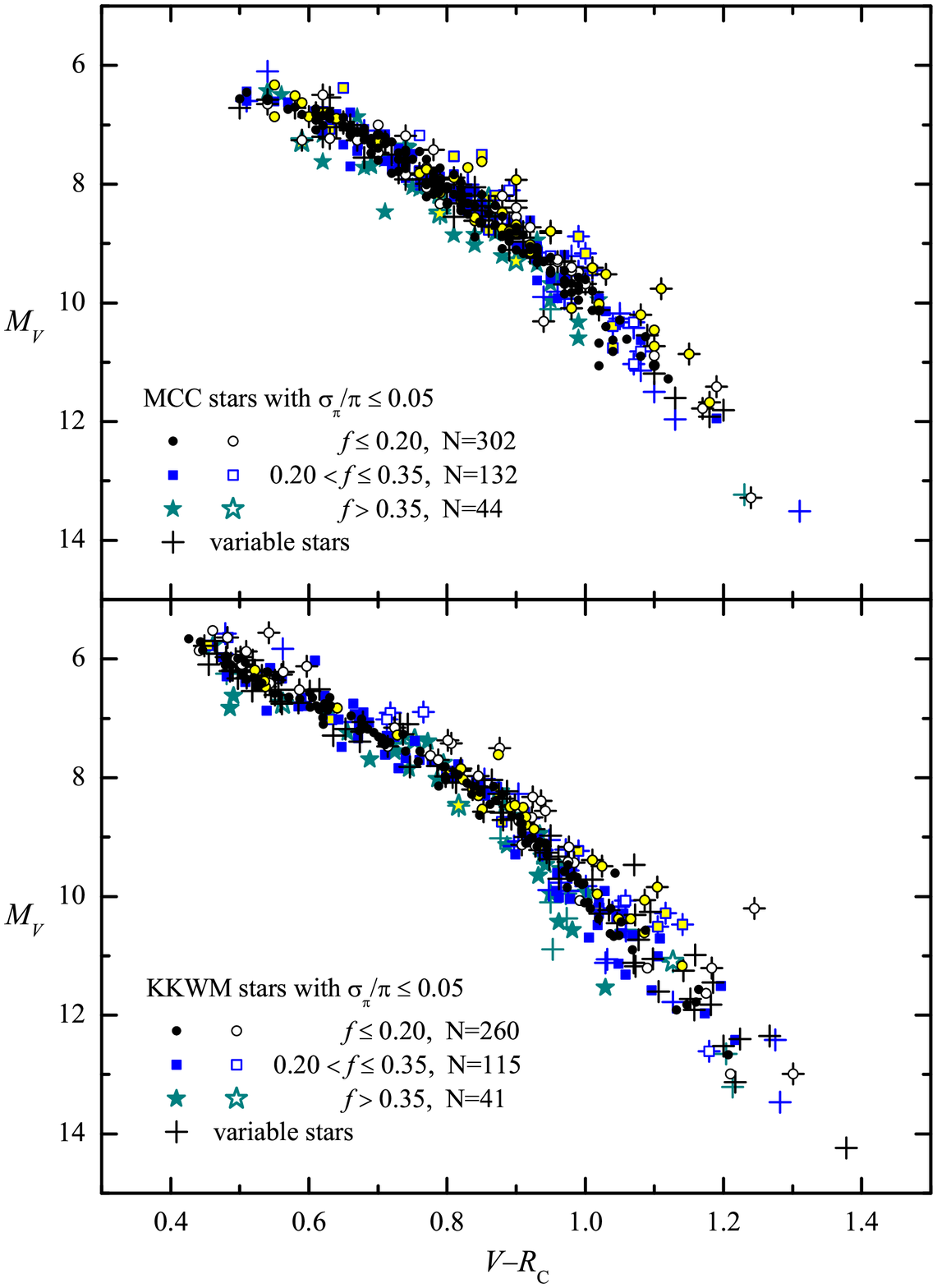,width=90mm,angle=0,clip=}}
\vskip-1mm
\captionb{3}{$M_V$, $V$--$R_{\rm C}$ diagram for the same
samples of K--M dwarfs as in Figure~2.}
\end{figure}

 \begin{figure}[t!]
\centerline{\psfig{figure=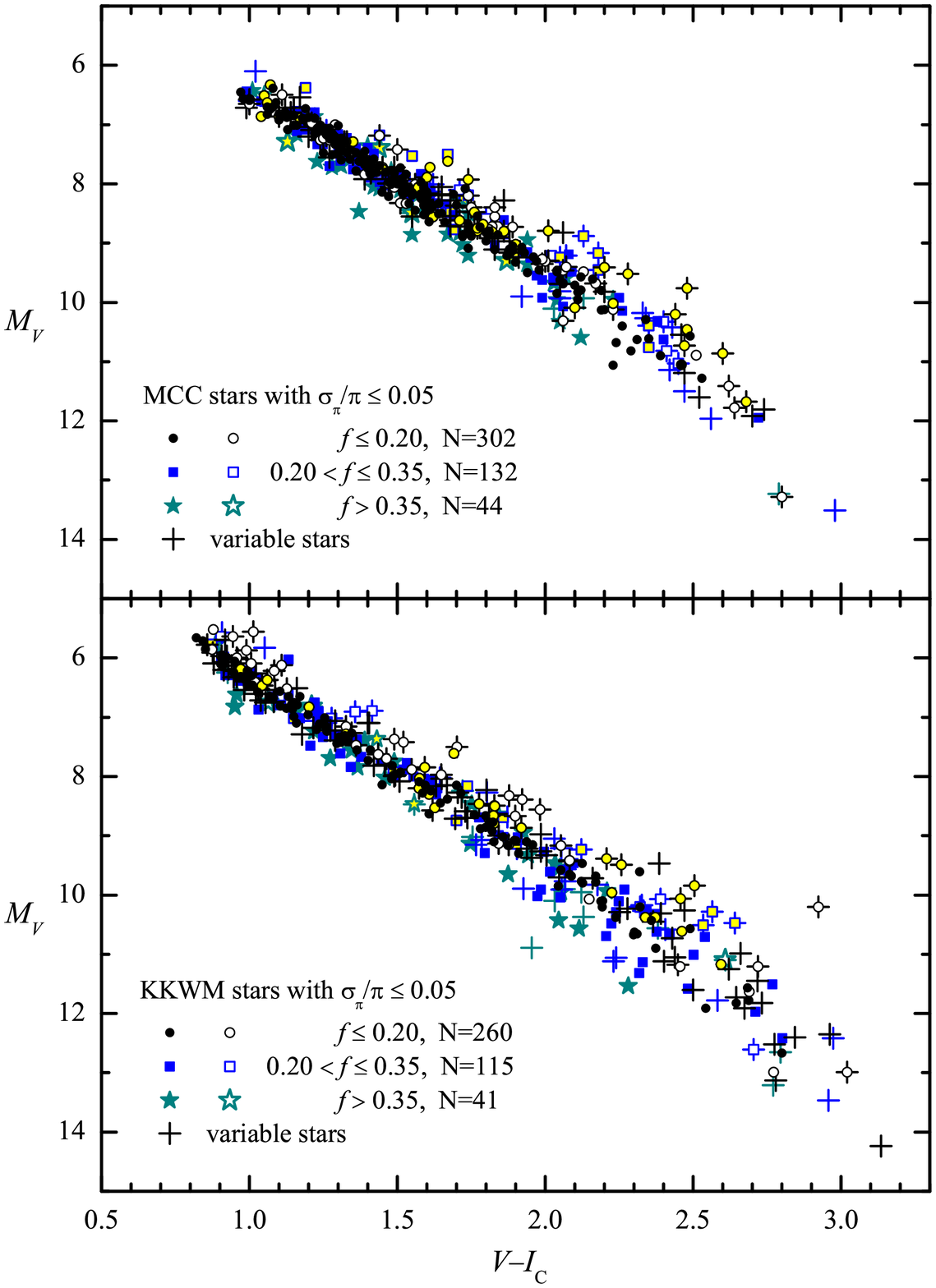,width=90mm,angle=0,clip=}}
\vskip-1mm
\captionb{4}{$M_V$, $V$--$I_{\rm C}$ diagram for the same
samples of K--M dwarfs as in Figure~2.}
\end{figure}

In Figures 2--4 we present ($M_V$, color) diagrams for the two
samples. To demonstrate what effect the addition of stars known or
suspected to be variables, binaries and belonging to presumably
metal-deficient population has on the width of the observational
sequence, we plotted all these groups of stars by different symbols.
We applied the so-called kinematical age parameter $f(U,V,W)$,
introduced by Grenon (1987),
\begin{equation}
           f = 1/C(a_1U^2+a_2V^2+a_3W^2)\,,
\end{equation}
to divide the stars into different populations: young- to
intermediate-age thin disk ($f\leq 0.20$), old thin disk
($0.20<f\leq0.35$) and thick disk ($f>0.35$). Here the space
velocity components $U$, $V$ and $W$ are computed with the values of
the Sun's motion relative the LSR
$(U_0,V_0,W_0)=(-10.0,12.2,7.2)$\,km\,s$^{-1}$ adopted from
Sch{\"o}nrich et al. (2010). The normalization constant $C$ and the
coefficients $a_i$ are adopted such that for stars on nearly
circular galactic orbits the parameter $f$ equals to orbital
eccentricity. As can be seen in the figures, the stars with the
thick disk kinematics are seen to lie below the sequence defined by
the rest of the stars, and this is an indication of their having
lower metallicity. The location of the sequence of stars with the
old-thin-disk kinematics is nearly identical to that of the younger
thin disk stars but differ in its larger (by 0.1 mag) scatter.

As one would expect, the known and suspected multiple stars tend to
lie above the sequence of their single counterparts at the same
color. The fraction of such stars, including those with the {\it
Hipparcos} multiplicity flags, amounts to 30\%. The spectroscopic
binaries (SB) and radial-velocity variables (SB candidates) comprise
13\% of the MCC sample for which our radial-velocity program is
nearing completion. Contrary to the general trend, noted in the
literature (e.g. Lada 2006), that the binary fraction is likely to
steadily decline with spectral type, we do not see from our
preliminary analysis of radial-velocity measurements a smaller
fraction of SB candidates among M-type dwarfs than among K-type
dwarfs (in both cases, around 10\%). Based on our radial-velocity
observations we report 24 new SB candidates among the MCC stars
considered, in addition to those for which radial-velocity
variability was noted or suspected in the literature.

The stars showing any signs of photometric variability, i.e. those
found in the GCVS database (Samus et al. 2007--2011) or flagged in
the {\it Hipparcos} catalog, constitute one third of the two samples
(in Figures 2--4 indicated by crosses).


\begin{center}
\vbox{\small \centerline{\parbox{120mm}{\baselineskip=9pt {\normbf
Table 1.}{ Ridge lines for young- to intermediate-kinematic-age
stars ($f\leq 0.20$): (1) MCC sample (160 stars), (2) KKWM sample
(105 stars). Given in the bottom line are the standard deviations of
single points from the ridge lines.}}}}
\smallskip
\vbox{ \footnotesize\tabcolsep=3pt
\begin{tabular}{rrrc|rrrrrc|rrrrr}
\hline
& & & & & & & & & & & & & &\\[-7pt]
 $B$--$V$ & $M_V^{\rm (1)}$ & $M_V^{\rm (2)}$ & & $V$--$R_{\rm C}$ & $M_V^{\rm (1)}$ & $M_V^{\rm (2)}$ & $M_{R_{\rm C}}^{\rm (1)}$ & $M_{R_{\rm C}}^{\rm (2)}$ & & $V$--$I_{\rm C}$ & $M_V^{\rm (1)}$ & $M_V^{\rm (2)}$ & $M_{I_{\rm C}}^{\rm (1)}$ & $M_{I_{\rm C}}^{\rm
 (2)}$\\[4pt]
\hline
& & & & & & & & & & & & & &\\[-8pt]
0.85 &      &  6.0 & & 0.45 &      &  5.8 &      &  5.4 & & 0.90 &      &  6.0 &      &  5.1 \\
0.90 &      &  6.2 & & 0.50 &      &  6.2 &      &  5.7 & & 1.00 &  6.5 &  6.4 &  5.4 &  5.4 \\
0.95 &  6.5 &  6.4 & & 0.55 &  6.6 &  6.5 &  6.1 &  5.9 & & 1.10 &  6.8 &  6.7 &  5.6 &  5.6 \\
1.00 &  6.7 &  6.6 & & 0.60 &  6.8 &  6.7 &  6.2 &  6.1 & & 1.20 &  7.1 &  7.0 &  5.9 &  5.8 \\
1.05 &  6.9 &  6.8 & & 0.65 &  7.1 &  7.0 &  6.4 &  6.3 & & 1.30 &  7.4 &  7.3 &  6.1 &  6.0 \\
1.10 &  7.1 &  6.9 & & 0.70 &  7.4 &  7.2 &  6.7 &  6.6 & & 1.40 &  7.7 &  7.6 &  6.3 &  6.2 \\
1.15 &  7.3 &  7.1 & & 0.75 &  7.7 &  7.6 &  6.9 &  6.8 & & 1.50 &  8.0 &  7.9 &  6.5 &  6.4 \\
1.20 &  7.4 &  7.3 & & 0.80 &  8.1 &  7.9 &  7.3 &  7.1 & & 1.60 &  8.3 &  8.2 &  6.7 &  6.6 \\
1.25 &  7.7 &  7.5 & & 0.85 &  8.5 &  8.3 &  7.6 &  7.5 & & 1.70 &  8.6 &  8.5 &  6.9 &  6.8 \\
1.30 &  7.9 &  7.7 & & 0.90 &  9.0 &  8.8 &  8.1 &  7.9 & & 1.80 &  8.9 &  8.8 &  7.1 &  7.0 \\
1.35 &  8.2 &  8.0 & & 0.95 &  9.5 &  9.3 &  8.5 &  8.4 & & 1.90 &  9.2 &  9.1 &  7.3 &  7.2 \\
1.40 &  8.5 &  8.3 & & 1.00 & 10.0 &  9.9 &  9.0 &  8.9 & & 2.00 &  9.5 &  9.4 &  7.5 &  7.4 \\
1.45 &  8.9 &  8.8 & & 1.05 & 10.5 & 10.5 &  9.5 &  9.4 & & 2.10 &  9.8 &  9.7 &  7.7 &  7.6 \\
1.50 &  9.6 &  9.7 & & 1.10 & 11.1 & 11.1 & 10.0 & 10.0 & & 2.20 & 10.1 & 10.0 &  7.9 &  7.8 \\
1.55 & 10.6 & 11.1 & & 1.15 &      & 11.8 &      & 10.6 & & 2.30 & 10.4 & 10.4 &  8.1 &  8.1 \\
1.60 &      & 12.8 & & 1.20 &      & 12.4 &      & 11.2 & & 2.40 & 10.8 & 10.7 &  8.3 &  8.3 \\
     &      &      & &      &      &      &      &      & & 2.50 & 11.1 & 11.1 &  8.5 &  8.6 \\
     &      &      & &      &      &      &      &      & & 2.60 &      & 11.5 &      &  8.9 \\
     &      &      & &      &      &      &      &      & & 2.70 &      & 12.0 &      &  9.3
     \\[1pt]
\hline
& & & & & & & & & & & & & &\\[-8pt]
s.d. & 0.22 & 0.32 &  &    & 0.17 & 0.17 & 0.17 & 0.17 &  &    &
0.18 & 0.19 & 0.18 & 0.19
\\[1pt]
\hline
\end{tabular}
}
\end{center}


\begin{wrapfigure}[21]{r}[0pt]{72mm}
\psfig{figure=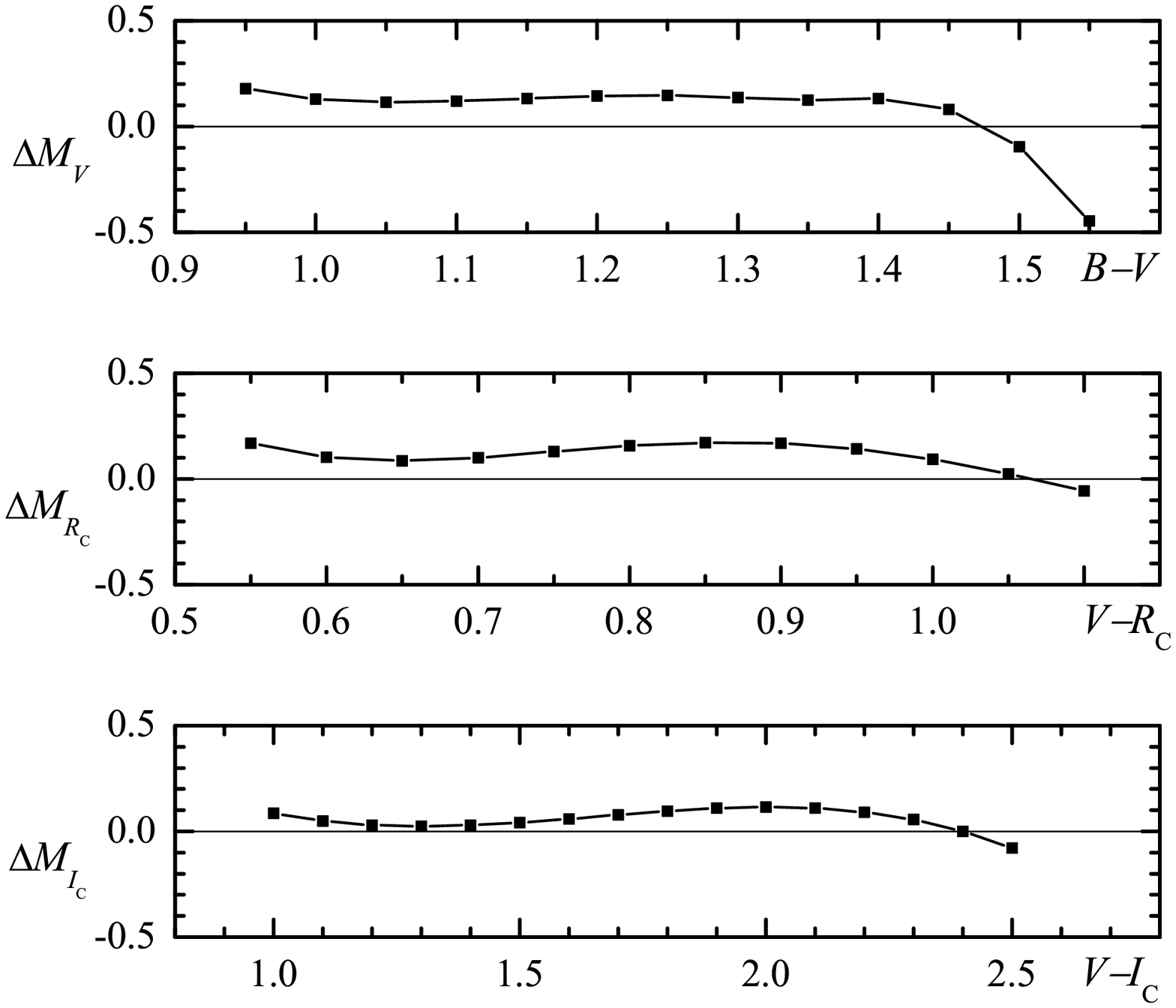,width=72mm,angle=0,clip=} \vspace{.2mm}
\vskip-5mm \captionb{5}{Differences in absolute magnitudes between
the ridge lines defined using the two samples, in the sense MCC
minus KKWM.}
\end{wrapfigure}

For comparison with theoretical isochrones, we removed stars which
may contribute increased scatter or broadening of the observational
sequence: known and probable multiple stars, variables, and stars
with $f$$>$\,$0.20$ which may belong to populations of subsolar
metallicity. After refining the MCC and KKWM samples in this way, we
are finally left with 160 and 105 stars in each sample,
respectively. The ridge lines of the stellar loci in the absolute
magnitude versus color planes, defined using each refined sample
separately, are tabulated in Table~1 (in the column headings,
superscripts (1) and (2) refer to MCC and KKWM, respectively). We
note that over most of the color range the ridge lines of the KKWM
sequences are systematically more luminous than those of the MCC
stars, with average offsets of 0.1 mag in $M_V$, $M_{R_{\rm C}}$ and
$M_{I_{\rm C}}$ (see Figure~5). Since systematic differences in
$BV(RI)_{\rm C}$ magnitudes between the MCC and KKWM data sets are
found to be of 0.02--0.03 mag (in the sense that KKWM magnitudes are
brighter), the much larger offsets in the color-absolute magnitude
space can be explained by a larger fraction of unrecognized binaries
left in the KKWM sample, which may lead to a shift of the ridge
lines toward brighter absolute magnitudes. Indeed, most of the KKWM
stars were not targets for our long-term radial-velocity program and
the fraction of known/suspected SB among them (7\%) is much less
significant than that among the MCC stars (13\%) covered entirely by
our program.

\sectionb{3}{COMPARISON WITH MODELS}

In Figure~6 we compare a set of solar composition isochrones to the
sequences of K--M dwarfs with $f\leq0.20$, cleaned of variable and
binary stars and SB candidates. The imposed cut in the kinematical
age parameter $f$ ensures that the stars belong to the young and
intermediate age population and are, on average, likely to be of
solar metallicity. We have tested 5 Gyr isochrones (the effect of
age is minor in the K--M dwarf range) of the following stellar
evolution models:
\begin{itemize}\addtolength{\itemsep}{-0.5\baselineskip}
    \item[(1)]Dartmouth (Dotter et al. 2008);
    \item[(2)]Padova (Marigo et al. 2008);
    \item[(3)]BaSTI (Cordier et al. 2007);
    \item[(4)]Victoria-Regina (VandenBerg et al. 2006);
    \item[(5)]Yonsei-Yale ${\rm Y}^2$
(Demarque et al. 2004) with the older (GDK, Green et al. 1987)
magnitude/color transformations;
    \item[(6)]Geneva (Lejeune \& Schaerer 2001);
    \item[(7)]Siess et al. (2000).
\end{itemize}

In the $M_V$,$B$--$V$ diagram of Figure~6, we also show the
empirical relation from the calibrations by Schmidt-Kaler (1982),
and, in the $M_V$,$V$--$I_{\rm C}$ diagram, the fiducial line of the
solar-metallicity open cluster M67, taken from Sandquist (2004) but
extended to the redder colors ($V$--$I_{\rm C}\geq1.6$) using
photometric data from the WEBDA Data Base. To aid visualization, we
have plotted in the figure the sequences of the MCC and KKWM stars
rather than their ridge lines. Note that photometry for 108 MCC
stars not common to the KKWM sample comes from the CNS3 database,
while for 52 stars common to both samples and for the rest 53 KKWM
stars is taken from KKWM. Therefore, plotted in the figure are
comparable numbers of stars representing the two sources of
$BV(RI)_{\rm C}$ photometry.

We conclude by Figure~6 that, with a few exceptions, most of the
isochrones provide a satisfactory fit to the dwarfs of spectral
types earlier than K7 ($B$--$V$$<$1.3), but fail to reproduce the
stellar locus of the less luminous dwarfs. In the region of M stars
($B$--$V$$\geq$1.4), some isochrones diverge from the main sequence
by more than 1 mag. Most appropriate are the Victoria-Regina
isochrones, which provide a good fit to the shape of the main
sequence down to the limit of their models (around M5), and the
BaSTI isochrones, matching the locus of K dwarfs well, especially in
the $M_V$,$B$--$V$ plane. The BaSTI colors $V$--$R_{\rm C}$ and
$V$--$I_{\rm C}$ at the reddest end (around K8--M0), however, do not
correctly match the data. In the $M_V$ versus $V$--$R_{\rm C}$ and
$V$--$I_{\rm C}$ planes, the Dartmouth models and isochrones by
Siess et al. (2000) also fit the data, but only down to spectral
type K8. There is also a general agreement between the sequence of
field K--M dwarfs and the fiducial line of M67.

\begin{figure}[hp!]
\centerline{\psfig{figure=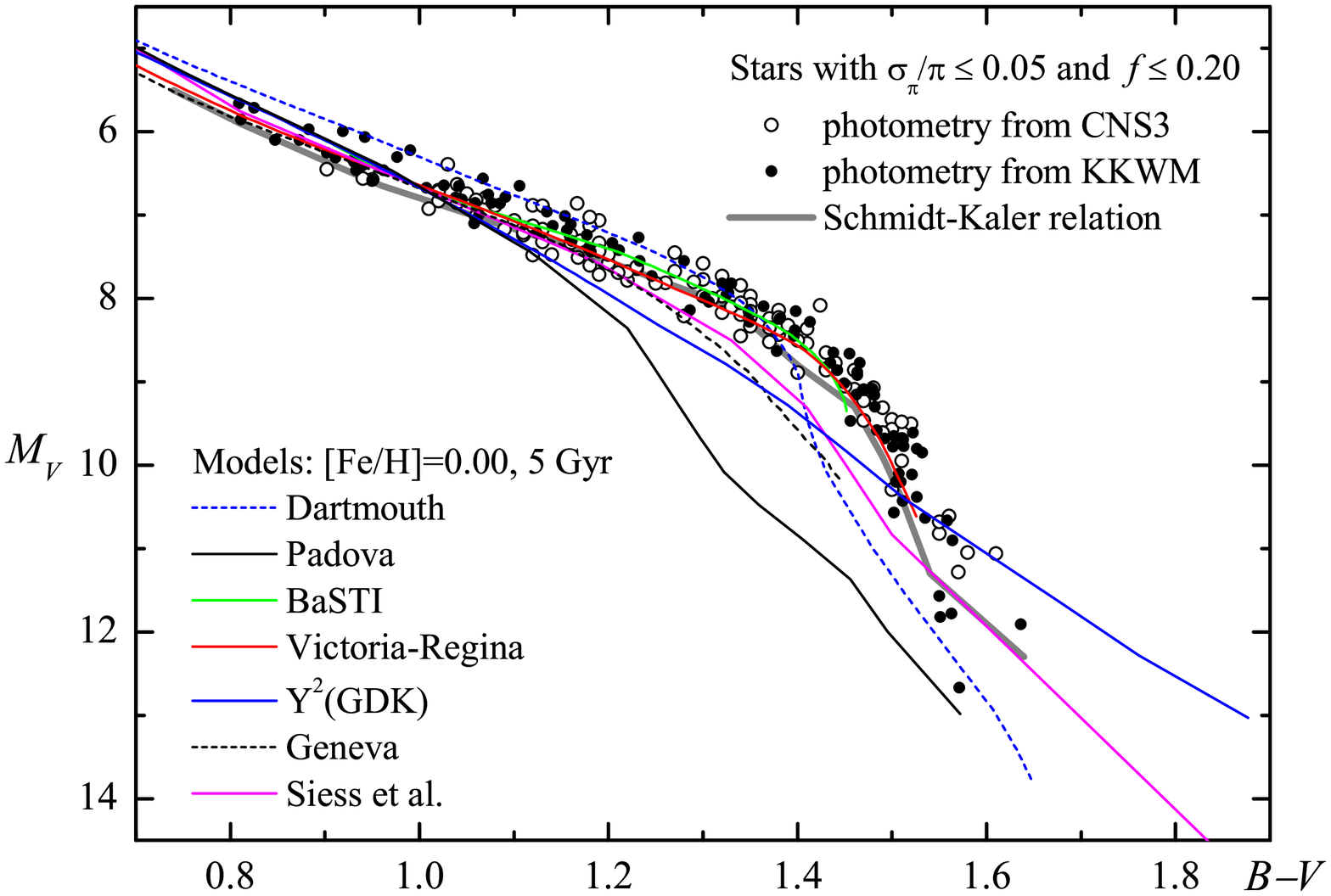,width=90mm,angle=0,clip=}}
\vskip3mm
\centerline{\psfig{figure=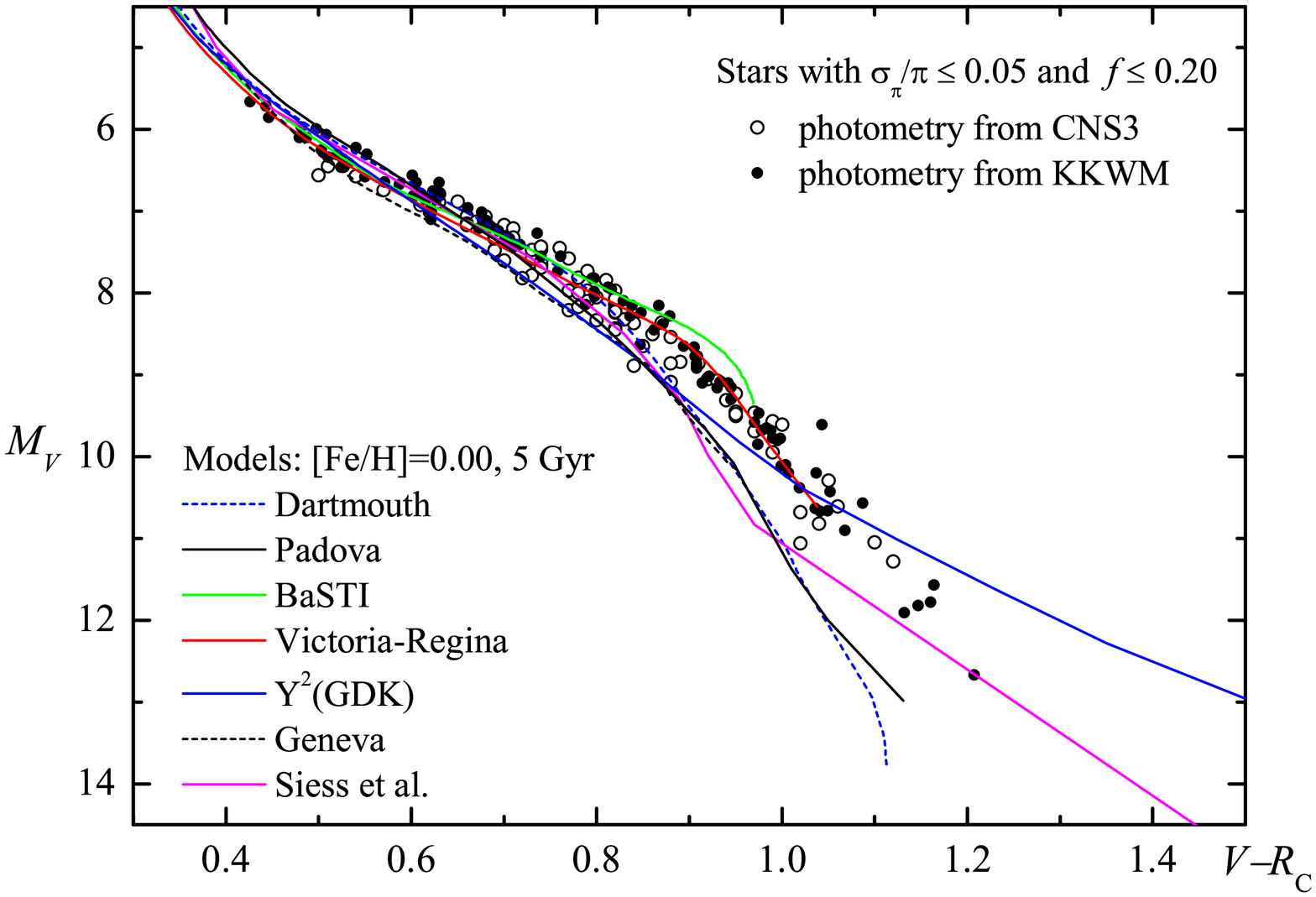,width=90mm,angle=0,clip=}}
\vskip3mm
\centerline{\psfig{figure=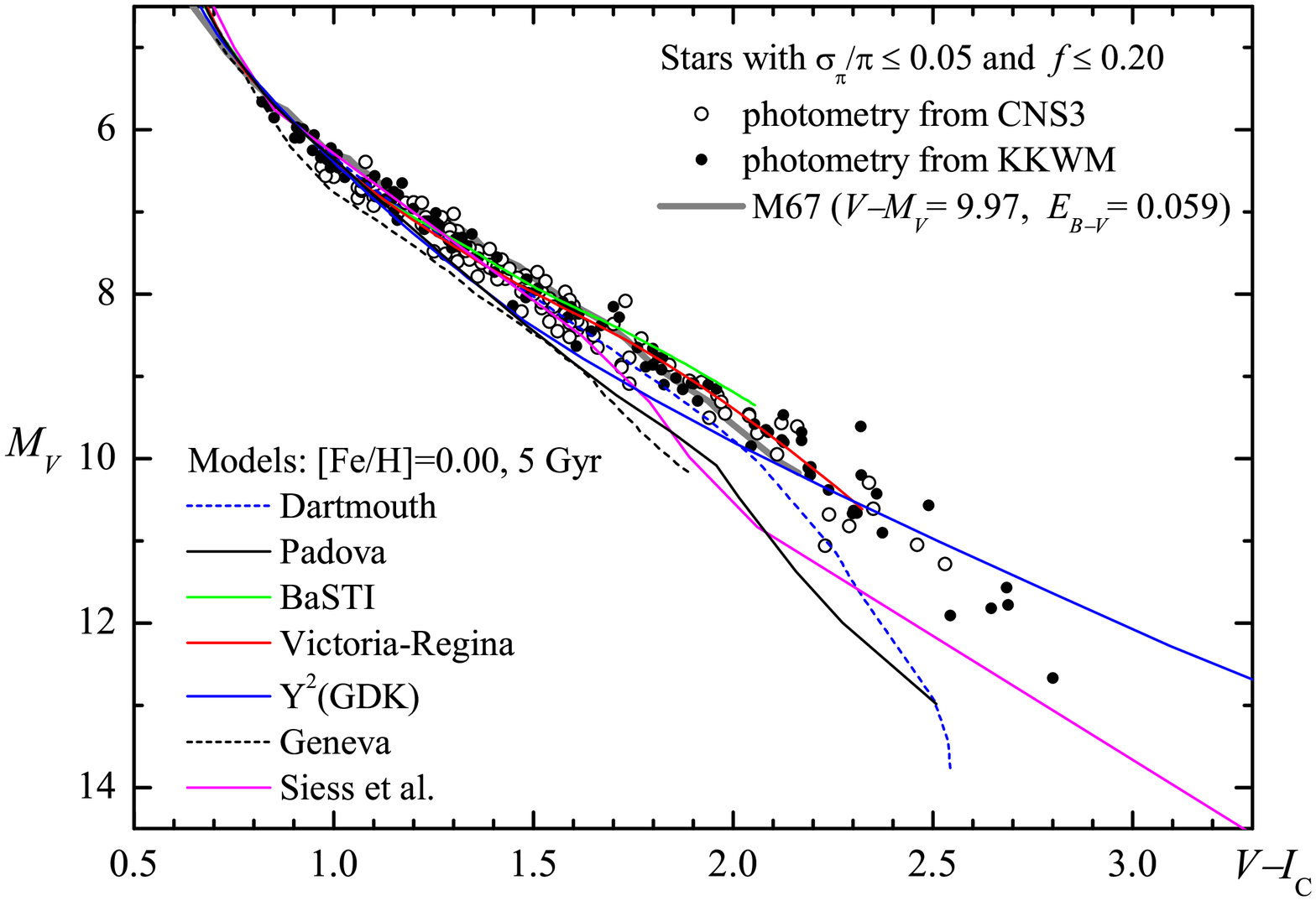,width=90mm,angle=0,clip=}}
\captionb{6}{Comparison of the observational ($M_V$,
color) sequences for K--M dwarfs with $f\leq 0.20$ to solar
metallicity theoretical isochrones of different models.}
\end{figure}

In summary, the samples of nearby K--M dwarfs show that there is a
clear problem in fitting most of the stellar models to low mass
stars. The Victoria-Regina isochrones (VandenBerg et al. 2006) with
empirically constrained color-temperature relations by VandenBerg \&
Clem (2003) appear to be the only models which are entirely
consistent with the observed loci of lower main-sequence stars in
the $BV(RI)_{\rm C}$ system.

\thanks{This research has made use of the SIMBAD data base and the VizieR
catalogue access tool, CDS, Strasbourg. The ARICNS data base,
Astronomisches Rechen-Institut, Heidelberg, and the WEBDA database,
Viena, have also been our data source. Direct access to the BaSTI,
Dartmouth, Geneva, Padova, Victoria-Regina and Yonsei-Yale databases
and the isochrone browse tools by Lionel Siess are greatly
acknowledged. The work was supported by the Research Council of
Lithuania under the grant No. MIP-132/2010.}

\References

\refb Bochanski J. J., Hawley S. L., Covey K. R. et al. 2010, AJ,
139, 2679

\refb Cordier D., Pietrinferni A., Cassisi S., Salaris M. 2007, AJ,
133, 468\\http://albione.oa-teramo.inaf.it/

\refb Demarque P., Woo J.-H., Kim Y.-Ch., Yi S. K. 2004, ApJS, 155,
667\\http://www.astro.yale.edu/demarque/yyiso.html

\refb Dotter~A., Chaboyer~B., Jevremovi{\'c}~D. et al. 2008, ApJS,
178, 89\\http://stellar.dartmouth.edu/~models/isolf.html

\refb Gliese W., Jahreiss H. 1991. Preliminary Version of the Third
Catalogue of Nearby Stars (CNS3); update at
http://www.ari.uni-heidelberg.de/datenbanken/aricns/

\refb Green E. M., Demarque P., King C. R. 1987, The Revised Yale
Isochrones and Luminosity Functions, New Haven, Yale Univ. Obs.
(GDK)

\refb Grenon M. 1987, Journ. Astron. \& Astroph., 8, 123

\refb Just A., Jahrei{\ss} H. 2008, Astron. Nachr., 329, 790

\refb Koen C., Kilkenny D., van Wyk F., Marang F. 2010, MNRAS, 403,
1949 (KKWM)

\refb Kotoneva E., Flynn C., Jimenez R. 2002, MNRAS, 335, 1147

\refb Lada C. J. 2006, ApJ, 640, L63

\refb Lejeune T., Schaerer D. 2001, A\&A, 366, 468\\
http://webast.ast.obs-mip.fr/equipe/stellar


\refb Marigo P., Girardi L., Bressan A. et al. 2008, A\&A, 482,
883\\
http://stev.oapd.inaf.it/cmd

\refb Samus N. N., Durlevich O. V., Kazarovets E. V. et al.
2007--2011, General Catalogue of Variable Stars (GCVS database,
Version 2011Jan), CDS B/gcvs

\refb Sandquist E. L. 2004, MNRAS, 347, 101

\refb Schmidt-Kaler T. 1982, in Landolt-B{\"o}rnstein New Series,
Group VI, vol. 2b, eds. K. Schaifers \& H. H. Voigt,
Springer-Verlag, Berlin

\refb Sch{\"o}nrich R., Binney J., Dehnen W. 2010, MNRAS, 403, 1829

\refb Siess L., Dufour E., Forestini M. 2000, A\&A, 358, 593\\
http://www.astro.ulb.ac.be/~siess/WWWTools/Isochrone

\refb Upgren A. R., Weis E. W. 1989, in {\it Star catalogues: a
centennial tribute to A. N. Vyssotsky}, eds. A. G. D. Philip \& A.
R. Upgren, L. Davis Press, Schenectady, p. 19

\refb Upgren A. R., Sperauskas J., Boyle R. P. 2002, Baltic
Astronomy, 11, 91

\refb VandenBerg D. A., Clem J. L. 2003, AJ, 126, 778

\refb VandenBerg D. A., Bergbusch P. A., Dowler P. D. 2006, ApJS,
162, 375;~http://\\
www1.cadc-ccda.hia-iha.nrc-cnrc.gc.ca/community/VictoriaReginaModels/

\refb van Leeuwen F. 2007, A\&A, 474, 653

\end{document}